\documentclass[10pt,conference]{IEEEtran}
\IEEEoverridecommandlockouts
\usepackage{amsmath}
\usepackage{amssymb,theorem}
\usepackage{graphics}
\usepackage{epsfig}
\usepackage[pdftex]{color}
\usepackage{multicol}

\long\def\comment#1{}

\newfont{\bbb}{msbm10 scaled 700}

\newfont{\bb}{msbm10 scaled 1100}
\newcommand{\CC}{\mbox{\bb C}}

\newcommand{\EE}{\mbox{\bb E}}

\newcommand{\hv}{{\bf h}}

\newcommand{\rv}{{\bf r}}

\newcommand{\uv}{{\bf u}}
\newcommand{\vv}{{\bf v}}
\newcommand{\wv}{{\bf w}}
\newcommand{\xv}{{\bf x}}
\newcommand{\yv}{{\bf y}}
\newcommand{\zv}{{\bf z}}

\newcommand{\Am}{{\bf A}}

\newcommand{\Dm}{{\bf D}}

\newcommand{\Hm}{{\bf H}}
\newcommand{\Id}{{\bf I}}

\newcommand{\Km}{{\bf K}}

\newcommand{\Pm}{{\bf P}}
\newcommand{\Qm}{{\bf Q}}

\newcommand{\Cc}{{\cal C}}

\newcommand{\Nc}{{\cal N}}

\newcommand{\Sigmam}{\hbox{\boldmath$\Sigma$}}

\newcommand{\SINR}{{\sf SINR}}

\newcommand{\herm}{{\sf H}}

\begin{document}

\title{On the Coexistence of Macrocell Spatial Multiplexing and Cognitive Femtocells}

\author{
\IEEEauthorblockN{Ansuman Adhikary}
\IEEEauthorblockA{
University of Southern California \\
Los Angeles, CA 90089, USA \\
adhikary@usc.edu}
\and
\IEEEauthorblockN{Giuseppe Caire}
\IEEEauthorblockA{
University of Southern California \\
Los Angeles, CA 90089, USA \\
caire@usc.edu}
\thanks{This research was supported in part by NSF Grant CCF-CIF 0917343 and by an
ETRI-USC Collaborative Research Project.}
}

\maketitle

\begin{abstract}
We study a two-tier macrocell/femtocell system where the macrocell
base station is equipped with multiple antennas and makes use of
multiuser MIMO (spatial multiplexing), and the femtocells are
``cognitive''. In particular, we assume that the femtocells
are aware of the locations of scheduled macrocell users
on every time-frequency slot,  so that they can make decisions on their transmission opportunities accordingly.
Femtocell base stations are also equipped with multiple
antennas. We propose a scheme where the macrocell downlink (macro-DL)
is aligned with the femtocells uplink (femto-UL)
and, Vice Versa, the macrocell uplink (macro-UL) is aligned with the femtocells downlink
femto-DL).  Using a simple ``interference temperature'' power control in the macro-DL/femto-UL direction, 
and exploiting uplink/downlink duality and the Yates, Foschini and Miljanic distributed power control algorithm
in the macro-UL/femto-DL direction, we can achieve an extremely attractive
macro/femto throughput tradeoff region in both directions. 
We investigate the impact of multiuser MIMO spatial multiplexing in the macrocell
under the proposed scheme, and find that large gains are achievable by letting the macrocell
schedule groups of co-located users, such that the number of femtocells affected 
by the interference temperature power constraint is small.
\end{abstract}

\section{Introduction}

It is widely recognized that {\em spatial reuse} is the single most valuable resource to dramatically increase the throughput of
wireless cellular networks.  However, deploying a very dense cellular infrastructure, with
base station (BS) density that grows {\em linearly} with the
user density,  is not viable for a number of obvious practical and economical reasons.
On the other hand, user-deployed WLANs (e.g., IEEE 802.11) achieve such dense spatial reuse
in the unlicensed band. This solution has the advantage of providing very high data rates for short-range,
mostly in-home, communication, but does not handle mobility as efficiently as cellular systems.
Therefore, {\em licensed} cellular systems are naturally evolving towards two-tier architectures, where
a large number of user-deployed {\em femtocells} operate under a common
macrocell ``umbrella'', that fills in the gaps of the small cells tier and supports mobility.
A large body of theoretical and standardization studies on this topic has been produced
in recent years (for a small sample, see \cite{rangan2010femto,
chandrasekhar2008femtocell,yeh2008wimax,xia2010open,dhillon2011modeling}).

Most existing works focus on continuous transmission
of macrocell user terminals (macro-UTs) and on the calculation of the pdf of the signal-to-interference plus noise (SINR)
at given receivers, where the tail of the SINR pdf is related to the probability of outage.
This approach disregards the fact that the forthcoming generation of cellular
systems (notable, 3GPP LTE and IEEE 802.16m)
is based on TDMA/OFDM, where dynamic scheduling is used in the macrocell tier for the the
downlink (macro-DL) and for the uplink (macro-UL). For a macro-BS equipped with $M$ transmit antennas and serving
up to $M$ macro-UTs on any time-frequency slot, the set of served macro-UTs (and therefore their location in the cell) may change
on a slot by slot basis. This gives a statistical multiplexing opportunity to the femtocell tier: in the macro-DL slots,
only the femtocells in the vicinity of a served macro-UT create significant interference to the macrocell tier;
 in the macro-UL slots, only the femtocells in the vicinity of an active macro-UT suffer from significant interference
 from the macrocell tier.

\begin{figure}[ht]
\centerline{\includegraphics[width=6cm,height=3cm]{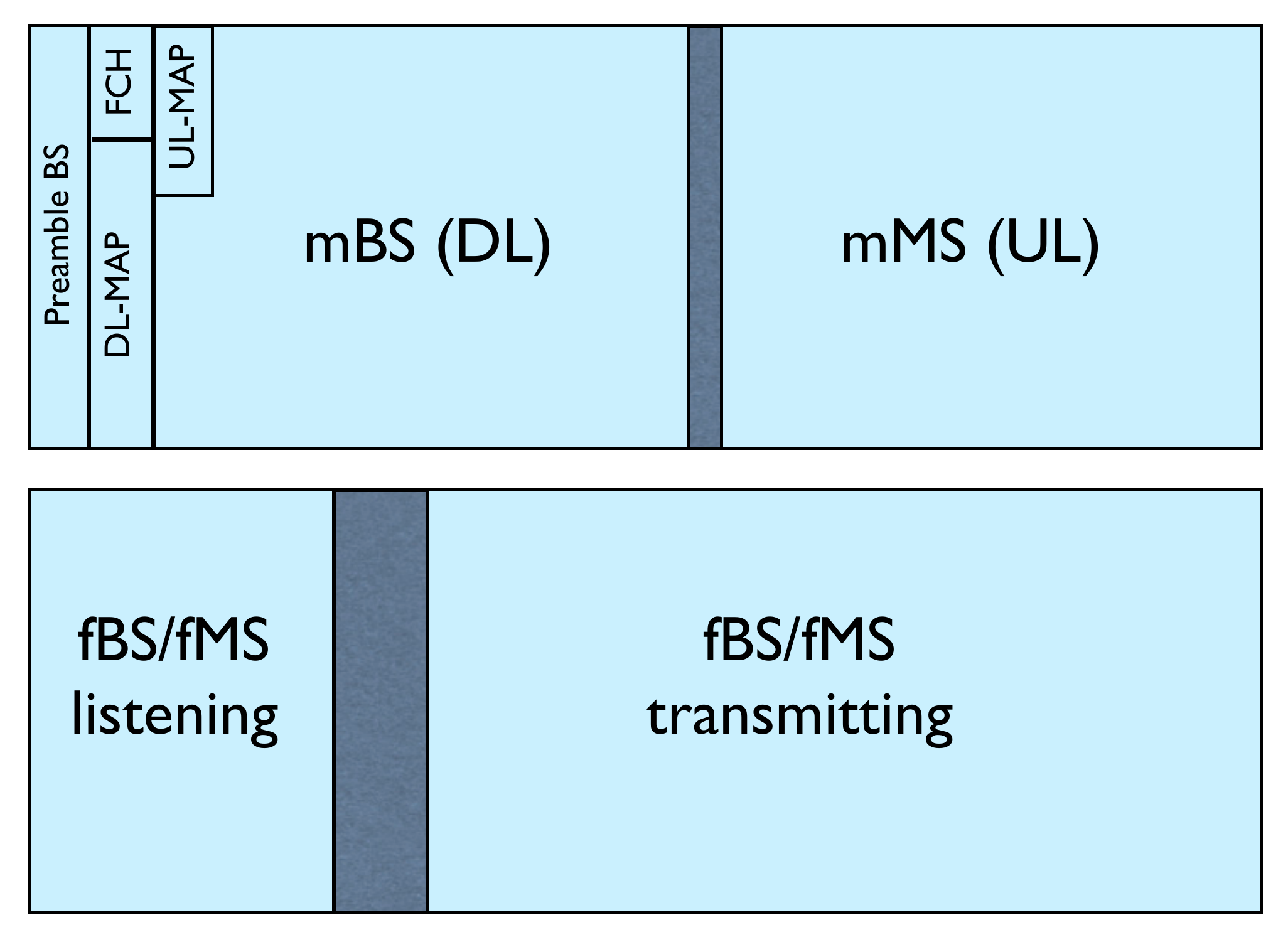}}
\caption{Frame structure for the proposed cognitive femtocell system.}
\label{femto-frame}
\end{figure}

In order to exploit the implicit statistical multiplexing due to the
macrocell dynamic scheduling, in \cite{caire-femtocell} we proposed
a ``cognitive'' approach to femtocells, where we assume that the
femto-BSs and the femto-UTs can decode the macro-BS allocation map
for both UL and DL. Fig.~\ref{femto-frame} shows a possible
arrangement of macro and femto frames allowing cognitive operations,
in the case where both tiers use time division duplexing (TDD). Assuming
that the positions of all terminals are known,~\footnote{Femtocells are deployed in fixed positions, that
can be made available through a database. Macro-UTs are mobile, but
their position changes sufficiently slow so that through GPS and
radio localization their position can be provided at a slow rate as
protocol side information by the macro-BS itself.}
the femtocells can regulate their transmit power in order to guarantee a given
``interference temperature'' to the macro-UTs. In
Section~\ref{sec:linear}, we review the details of a scheme proposed
in \cite{caire-femtocell}, based on linear beamforming and UL/DL
duality, and we extend it to the case of multi-antenna macro-BS.
Then, in Section~\ref{sec:coexistence} we discuss the coexistence of
the multiuser MIMO (MU-MIMO) spatial multiplexing in the macrocell
tier with the cognitive femtocells. In fact, it is intuitively clear
that there is a tradeoff between the macrocell and  the aggregate
femtocell throughput: if the macrocell serves many macro-UT using
spatial multiplexing, correspondingly many femtocells have to turn
down their transmit power because of the interference temperature
requirement, and therefore the femtocell throughput is decreased. In
contrast, if the macro-BS serves only one macro-UT at each
time-frequency slot, only a few femtocells are affected by the power
control requirement but the macrocell tier does not exploit the full
multiplexing gain and its throughput is decreased. We shall see that
this problem is alleviated by scheduling approximately co-located
groups of macro-UTs.

\section{SIMO/MISO interference channel} \label{sec:linear}

We consider a single macro-BS with $M$ antennas, serving $K \leq M$
macro-UTs. In the same coverage area, a set of  femtocells
share the same frequency band. Both tiers operate in TDD. The
channel gains are formed by two components: a pathloss factor
constant in time (over a large number of time-frequency slots) and frequency-flat,
and a time-frequency selective small-scale fading that changes
independently on a slot by slot basis.  For simplicity, we focus
here on a single subcarrier. By symmetry of the fading distribution,
our results extend directly to an OFDM system with independent
scheduling on each subcarrier.~\footnote{In general, a slight
improvement can be obtained by scheduling jointly across the
subcarriers. However, the improvement due to this more sophisticated
multiuser scheduling across frequencies is marginal with respect to
the system throughput achieved by the proposed scheme even with the
suboptimal per-subcarrier scheduling.}

\begin{figure}[ht]
\centerline{\includegraphics[width=8cm,height=5cm]{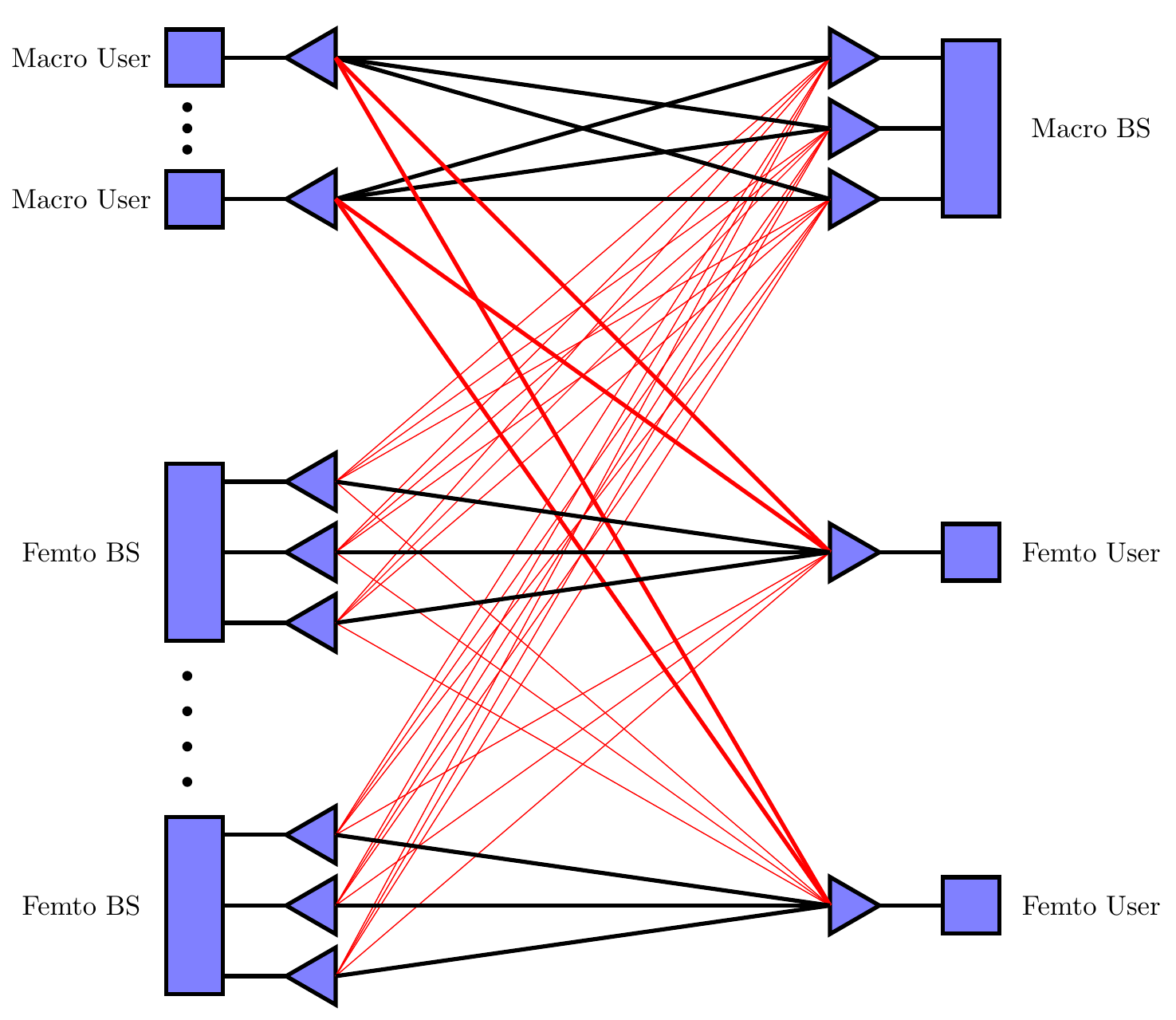}}
\caption{Interference channel model for the two-tier
network with multiple antennas at the femto-BSs and macro-BS.
The thick red lines indicate the links that determine the femto user
transmit power, via the interference temperature power control.}
\label{fig:mimo-duality}
\end{figure}

In the proposed scheme we have two types of slots: macro-DL/femto-UL and
macro-UL/femto-DL (see Fig.~\ref{fig:mimo-duality}).
The femtocells operate in TDMA.
Therefore, the number of femto-UTs actually present in
each femtocell is irrelevant, since only one of them is active at any given slot and, for the sake of simplicity,
it is sufficient to consider a single femto-UT per femtocell.
Notice that the femtocells form a SIMO/MISO interference channel,
coupled with the vector broadcast (DL) and multiaccess (UL) channel corresponding to the macrocell.

{\bf Macro-DL/Femto-UL slot:}
Macro- and femto-UTs are equipped with a single antenna. The received signal at the $k$-th macro-UT is given in general by
\begin{align} \label{eq:simo-hybrid}
y_k = & \; \sqrt{g(k,0}) \hv_{{\rm mc},k}^{\herm}   \sum_{i=1}^K \vv_i x_{{\rm mc},i} + \sum_{f \in \Cc} \sqrt{g(k,f)} h_{k,f} x_f + z_k
\end{align}
where $\hv_{{\rm mc},i} \in \CC^{M \times 1}$ is the
channel vector from the macro-BS antenna array to macro-UT
$i$, $\vv_i$ is the corresponding macro-BS beamforming vector, $\Cc$ denotes the set of all femtocells,
$h_{f,k}$ is the scalar small-scale fading coefficient from femto-UT $f$ to the macro-UT $k$, and $z_k \sim \Cc\Nc(0,1)$ is AWGN.
The coefficients $g(a,b)$ indicate pathloss between points $a$ and $b$, as detailed in Section \ref{sec:coexistence}.
The macro-BS is located at 0 (the origin of the cell coordinate system).
The macro-BS calculates the beamforming vectors as functions of the matrix
\begin{equation}
\Hm_{\rm mc} = [\hv_{{\rm mc},1},\ldots,\hv_{{\rm mc},K}],
\end{equation}
formed by the $K$ active macro-UTs, enumerated without loss of generality from 1 to $K$.
This can be obtained either by TDD reciprocity (open-loop) or by explicit channel state
feedback \cite{caire2010multiuser}. In particular, here we consider Linear Zero-Forcing Beamforming (LZFB),
such that  $\vv_i$ is given by the $i$-th  column of the Moore-Penrose pseudo-inverse of $\Hm_{\rm mc}$ normalized to have unit norm. Hence,
$\hv_{{\rm mc},k}^{\herm} \vv_i = 0$ for all $i \neq k$. The macro-BS is subject to a total transmit power equal to $P_0$,
equally allocated over the $K$ DL data symbols $x_{{\rm mc},i}$.

The femto-UT in the $f$-th femtocell transmits with power $\EE[|x_f|^2] = P_f$,  regulated
such that the interference caused at all active macro-UTs users
is less than the target interference temperature $\kappa$. Hence, we have
\begin{equation} \label{eq:femto-power-control}
P_f = \min \left \{ \frac{\kappa}{ \max_{k \in \{1,\ldots,K\}} g(k,f)}, P_1 \right \},
\end{equation}
where $P_1$ is the peak femtocell power.
The SINR for macro-UT $k$ is given by
\begin{eqnarray} \label{eq:sinr-macro}
\SINR_k^{\rm mc-DL} &=& \frac{g(k,0)|\hv_{{\rm mc},k}^{\herm} \vv_k|^2\frac{P_0}{K}}{1
+ \sum_{f \in \Cc} g(k,f) |h_{k,f}|^2 P_f}
\end{eqnarray}
and the corresponding instantaneous rate for macro-UT $k$ on the
current slot is given by $R_k = \log(1 + \SINR_k^{\rm mc-DL})$.
For simplicity, we assume that the macro-BS schedules at each slot
$K$ out of $U \gg K$ macro-UTs, picked at random with equal
probability. Hence, by averaging over the fading realization and the
${U \choose K}$ sets of macro-UTs, we obtain the sum-throughput of
the macrocell tier in the DL. When $K = M$ and the fading is
Rayleigh i.i.d.,  using the results in \cite{kim2011transmission},
this can be given in closed form for fixed $U$  macro-UT positions.

The femto-BS receivers have $L$ antennas each, and make use of
linear MMSE detection, which maximizes SINR over all linear receivers.
The received signal vector at a femto-BS $f$ is given by
\begin{align} \label{eq:simo-hybrid-femto}
\yv_f = & \; \sum_{j \in \Cc} \sqrt{g(f,j)} \hv_{f,j} x_{j} + \sqrt{g(f,0)} \Hm_{f,0}^{\herm}  \sum_{k=1}^K \vv_k x_{{\rm mc},k} + \zv_f
\end{align}
The linear MMSE receive vector for estimating the desired symbol $x_f$ from $\yv_f$ is given by
$\uv_f = \alpha_f \Sigmam_f^{-1} \hv_{f,f}$ where $\alpha_f > 0$ is chosen such that $\| \uv_f\| = 1$, and
$\Sigmam_f$ is the interference-plus-noise covariance matrix in (\ref{eq:simo-hybrid-femto}), given by
\begin{align} \label{sigmaf}
\Sigmam_f = & \; \Id +  \sum_{j \in \Cc:j \neq f} g(f,j) \hv_{f,j} \hv_{f,j}^{\herm} P_{j} \nonumber \\
& \; + g(f,0) \frac{P_0}{K} \Hm_{f,0}^{\herm}  \left (  \sum_{k=1}^K \vv_k \vv_k^{\herm} \right ) \Hm_{f,0} .
\end{align}
The receiver forms the scalar observation $\widehat{y}_f = \uv_f^{\herm} \yv_f$, and the corresponding SINR is given by
\begin{equation} \label{eq:sinr-femto}
\SINR_f^{\rm fc-UL} = P_{f} \hv_{f,f}^{\herm} \Sigmam_{f}^{-1} \hv_{f,f}.
\end{equation}
Similarly to what argued before,  the instantaneous rate of femtocell $f$ is  given by $R_{f} = \log (1 + {\rm SINR}_f^{\rm fc-UL})$.
By summing over all the femtocells and averaging over the fading and the ${U \choose K}$ active macro-UTs sets (notice that they have an influence through
the femtocell transmit powers $P_f$), we obtain the sum-throughput of the femtocell tier in the UL.

{\bf Macro-UL/Femto-DL:}
On the macro-UL/femto-DL slot, each femtocell has multiple antennas
whereas the macro-UTs have single antenna each. Insisting on
linear beamforming strategies, each femto-BS sends the
$L$-dimensional signal vector $\xv_f = \wv_f s_f$, where $\wv_f$ denotes the transmit
beamforming vector and $s_f$ is the corresponding (coded) data symbol
for its own intended femto-UT.

The received signal at the macro-BS  is given by
\begin{align} \label{eq:miso-hybrid}
\yv = & \; \sum_{k=1}^K \sqrt{g(k,0)} \hv_{{\rm mc},k} x_{{\rm mc},k} + \sum_{f \in \Cc} \sqrt{g(f,0)} \Hm_{f,0} \wv_f s_f + \zv
\end{align}
The BS forms the scalar observation $\widehat{y}_k = \rv_k^{\herm}
\yv_k$ for detecting $x_k$, where $\rv_k$ is the
receive beamforming vector for macro-UT $k$.
%

The received signal at the femto user in femtocell $f$ is given by
\begin{align} \label{eq:miso-hybrid-femto}
y_f = & \; \sum_{j \in \Cc} \sqrt{g(j,f)} \hv_{j,f}^{\herm} \wv_{j} s_{j}
+ \sum_{k=1}^K \sqrt{g(k,f)} h_{k,f}^{*} x_{{\rm mc},k} + z_f.
\end{align}
Calculating the instantaneous SINRs $\SINR_k^{\rm mc-UL}$ and
$\SINR_f^{{\rm fc-DL}}$ from (\ref{eq:miso-hybrid}) and  from
(\ref{eq:miso-hybrid-femto}), respectively, is straightforward.
In particular, from UL/DL duality (see \cite{viswanath2003sum}), extended to
the MISO/SIMO interference channel in \cite{negro2010beamforming}, 
we have that by letting  $\wv_f = \uv_f$ and $\rv_k = \vv_k$,
it is possible to achieve   $\SINR_k^{\rm mc-UL} = \SINR_k^{\rm mc-DL}$ and
$\SINR_f^{\rm fc-DL}  = \SINR_f^{\rm fc-UL}$ while preserving the total sum power, i.e., with
\begin{equation} \label{sum-power-equality}
\sum_{f \in \Cc} Q_f + \sum_{k=1}^K Q_{{\rm mc},k} = \sum_{f \in \Cc} P_f + P_0,
\end{equation}
where $Q_{{\rm mc},k}$ denotes the transmit power of macro-UT $k$ and $Q_f$ denotes
the transmit power of the femto-BS $f$. The power allocation across the macro-UTs and the femto-BSs
depends on the realization of the path losses and small scale fading components (which determine the
beamforming vectors).

\subsection{Implementation issues}

In order to calculate the beamforming vectors $\uv_f$,
using the matrix inversion lemma we can write $\uv_f = \beta_f \Km_f^{-1} \hv_{f,f}$,
where $\beta_f > 0$ is another
normalizing proportionality constant and $\Km_f$ is the received signal covariance matrix given by
\begin{equation}
\Km_f = \Sigmam_f  + g(f,f) \hv_{f,f} \hv_{f,f}^\herm P_f
\end{equation}
Hence, the MMSE beamforming vectors can be conveniently calculated by using a sample covariance
estimate of $\Km_f$, from the whole received femto-UL slot, and an estimate of the desired signal channel $\hv_{f,f}$
obtained by using UL pilots symbols, as in standard
coherent detection for MIMO channels (e.g., currently implemented in IEEE 802.11n).

The other practical implementation problem of the proposed scheme consists of calculating
the transmit powers $Q_{{\rm mc},k}$ and $Q_f$ in the macro-UL/femto-DL slot, for fixed
unit-norm beamforming vectors $\wv_f = \uv_f$ and $\rv_k = \vv_k$.
We propose to use the well-known Yates-Foschini-Miljanic distributed power allocation
algorithm (see \cite{yates1995framework,foschini1993simple}), that is guaranteed to converge to the solution.

For all femtocells $f$, fix the target DL SINR $\gamma_f^{\rm fc-DL} =  \SINR_f^{\rm fc-UL}$. For
all active users $k$ fix the target UL SINR $\gamma_k^{\rm mc-UL} = \SINR_k^{\rm mc-DL}$.
Let $\SINR_f^{\rm fc-DL}(\{ Q_f\}, \{ Q_{{\rm mc},k}\}, \{\uv_f\})$
denote the femtocell $f$
DL SINR for fixed beamforming vectors and transmit powers, and let
$\SINR_k^{\rm mc-UL}(\{ Q_f\}, \{ Q_{{\rm mc},k}\}, \{\uv_f\}, \{\vv_k\})$
denote the macro-UT $k$ UL SINR for fixed beamforming vectors and transmit powers.
Then, the iterative distributed power control algorithm, in our case,  is given by:
\begin{enumerate}
\item {\em Initialization:} let $n = 0$ and let $Q_{{\rm mc},k}^{(0)} = P_0/K$,
$Q_f^{(0)} = P_f$ for $k  = 1, \ldots, K$ and all $f \in \Cc$.
\item {\em Iterations:} for $n = 1, 2, 3, \ldots$ do
\end{enumerate}
\begin{eqnarray}
Q_{{\rm mc},k}^{(n)} &=& \frac{ Q_{{\rm mc},k}^{(n-1)}  \gamma_k^{\rm mc-UL}}{\SINR_k^{\rm mc-UL}(\{ Q_f^{(n-1)}\}, \{ Q_{{\rm mc},k}^{(n-1)}\}, \{\uv_f\}, \{\vv_k\})}
\nonumber \\
Q_f^{(n)} &=&  \frac{Q_f^{(n-1)} \gamma_f^{\rm fc-DL}}{\SINR_f^{\rm fc-DL}(\{ Q_f^{(n-1)}\}, \{ Q_{{\rm mc},k}^{(n-1)}\}, \{\uv_f\})}.  \nonumber \\
& &  \label{update-Q}
\end{eqnarray}
In order to implement this scheme, a sequence of adjacent slots should be allocated to the same group
of $K$ macro-UTs, and at each slot the receivers measure their SINR and report their measurements to the transmitters
such that the power values can be updated according to (\ref{update-Q}).

\section{Numerical results} \label{sec:coexistence}

In line with \cite{caire-femtocell}, we consider a unit-side square
cell $[-1/2,1/2] \times [-1/2,1/2]$, with the macro-BS located
at the origin 0, and $F^2$ femtocells are centered at points of
coordinates $\left(\frac{2i-F+1}{2F},\frac{2j-F+1}{2F}\right)$, for
$i,j = 0,\ldots,F-1$. Femtocells are disk-shaped with radius $r_{\rm
fc}$, shielded from the outdoor environment by walls. For two points
$a,b$, the distance dependent path loss component is given by
\begin{equation}
g(a,b) = \frac{w^{n(a,b)}}{1 + (d(a,b)/\delta)^{\alpha}}
\end{equation}
where
$d(a,b)$ denotes the distance between $a$ and $b$ modulo 
the centered unit square (torus topology); $n(a,b)$ counts the number of walls between points 
$a$ and $b$ (i.e., $n(a,b) = 0$ if both $a$ and $b$ are outdoor or they are in the same femtocell, 
$n(a,b) = 1$ if either $a$ or $b$ is indoor (inside a femtocell), 
and $n(a,b) = 2$ if $a$ and $b$ are in different femtocells); 
$w$ is the wall absorption factor; $\delta$ is the``3 dB'' pathloss distance; 
$\alpha$ is the outdoor pathloss exponent.

\begin{table}
\centering \caption{Simulation Parameters}
\begin{tabular}{|c|c|c|}
    \hline Parameter & Notation & Value\\
    \hline Macro Cell Side Length & $L$ & 1000 m\\
    \hline Path Loss Parameter & $\delta$ & 50 m\\
    \hline FC Radius & $r_0$ & 10 m\\
    \hline Distance between two FC & $l$ & 40 m\\
    \hline Path Loss Exponent & $\alpha$ & 3.5\\
    \hline Wall Partition Loss & $w$ & 5 dB\\
    \hline Min SNR at cell edge & ${\rm SNR}_{\rm min}$ & 10 dB\\
    \hline Number of antennas at macro-BS & $M$ & 8\\
    \hline Number of antennas at femto-BS & $L$ & 5\\
    \hline
\end{tabular}
\label{param-sim-list}
\end{table}

\begin{figure}
  \centering
  \includegraphics[width=8cm,height=6cm]{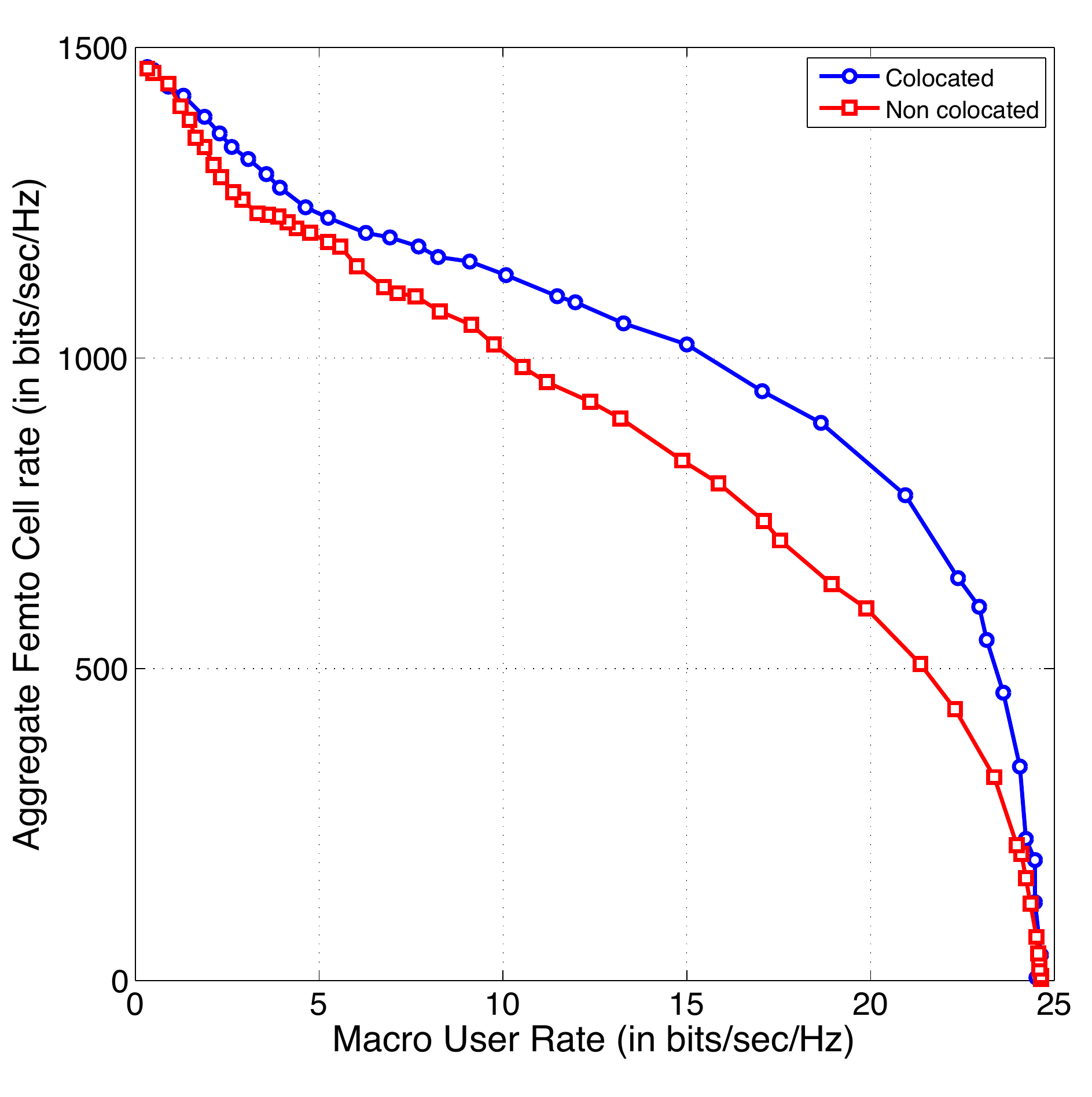}\\
  \caption{Throughput tradeoff region comparison (Pareto boundary) between colocated and non colocated users in 
  macro-DL/femto-UL. 
  }\label{fig:compare-tradeoff}
\end{figure}

We fix the macro-BS power $P_0$ such that the received SNR (without
interference) for a macro-UT at the cell edge is 10 dB. 
By varying the value of the interference power temperature $\kappa$ and 
letting $P_1 = 30$ dB,\footnote{Assuming a peak rate in a femtocell in isolation is 10
bit/s/Hz, corresponding to uncoded 1024 QAM, we have SNR $= 10
\log_{10}( 2^{10} - 1) = 10*\log_{10}(1023) \approx 30 $dB.} 
we obtain the Pareto boundary of the throughput tradeoff region achievable 
with the proposed scheme in the macro-DL/femto-UL slot, as described before. 
The tradeoff region for the dual channel, i.e., macro UL/femto DL is obtained by 
using the same beamforming vectors and the iterative power control algorithm. We
distinguish between the cases of colocated and non-colocated macro-UTs. In the first case, 
we macro-BS schedules $K$ users roughly located in the same position of the cell, such that they are separated enough
to have independent small-scale fading, but they have the same pathloss with respect to the macro-BS and all the femtocells. 
In the second case, the $K$ macro-UT positions are independently selected with uniform probability over the cell. 
Fig.~\ref{fig:compare-tradeoff} shows the comparison of the Pareto boundaries of the tradeoff regions
for the colocated and non-colocated case (supremizing over $K$), showing a clear advantage for the colocated case,
wich is possible when the macro-UT density is large enough so that $K$ approximately colocated users can be found. 
In all the results, femtocells are assumed to be ``open access'', therefore, macro-UTs are located only outdoor since a macro-UT inside a femtocell
would be automatically ``swallowed'' by the femtocell, and served as a femtocell user. 
We already showed in \cite{caire-femtocell} that the impact of closed-access femtocells is minimal, 
in contrast to  what observed for conventional ``legacy'' systems, thanks to the proposed cognitive scheme and the 
interference temperature power control. 

\begin{figure}
  \centering
  \includegraphics[width=8cm,height=6cm]{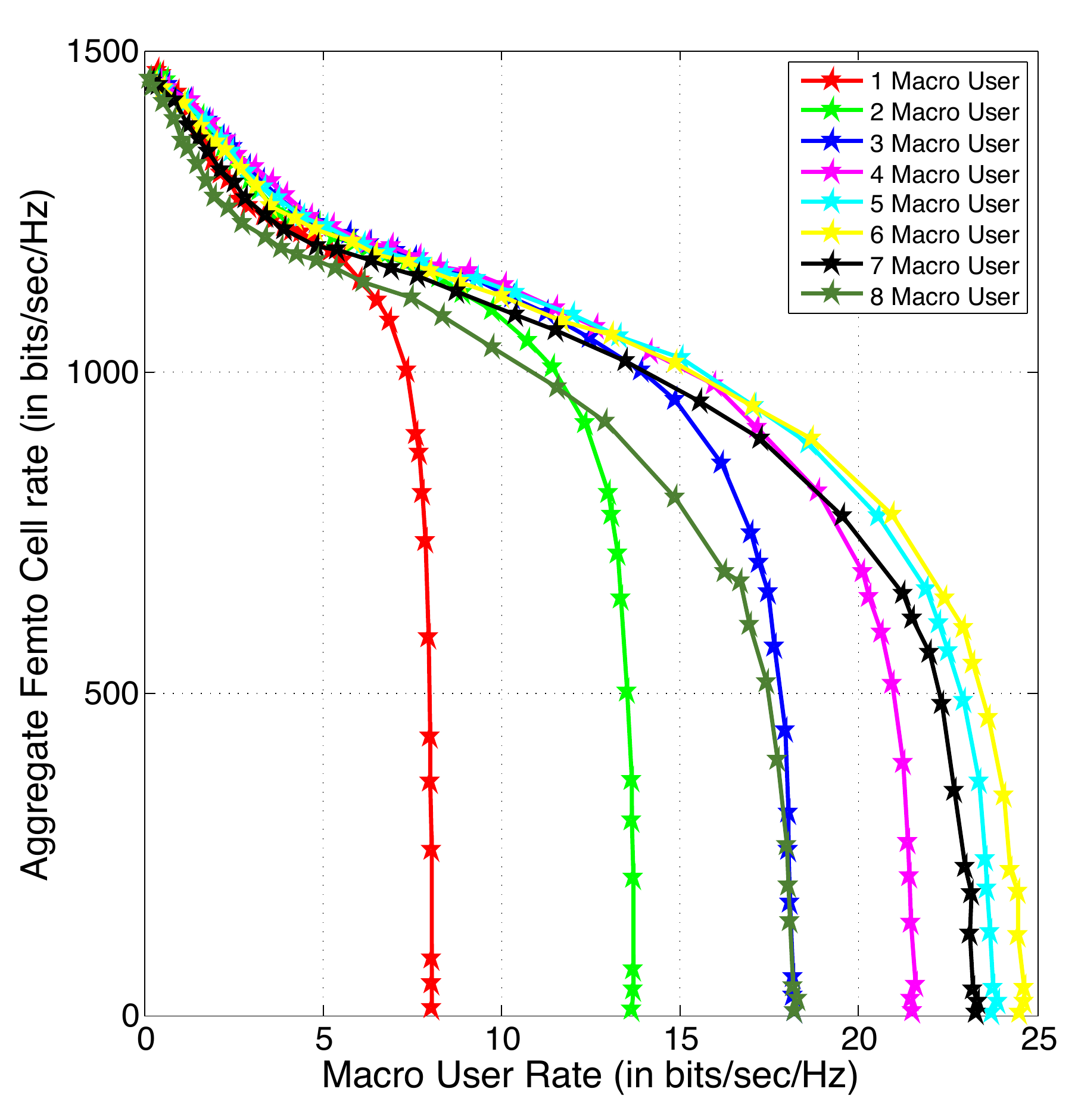}\\
  \caption{Throughput tradeoff region comparison for colocated users in macro-DL/femto-UL. 
  The different colors indicate the tradeoff region for different $K$.}\label{fig:colocated}
\end{figure}

\begin{figure}
  \centering
  \includegraphics[width=8cm,height=6cm]{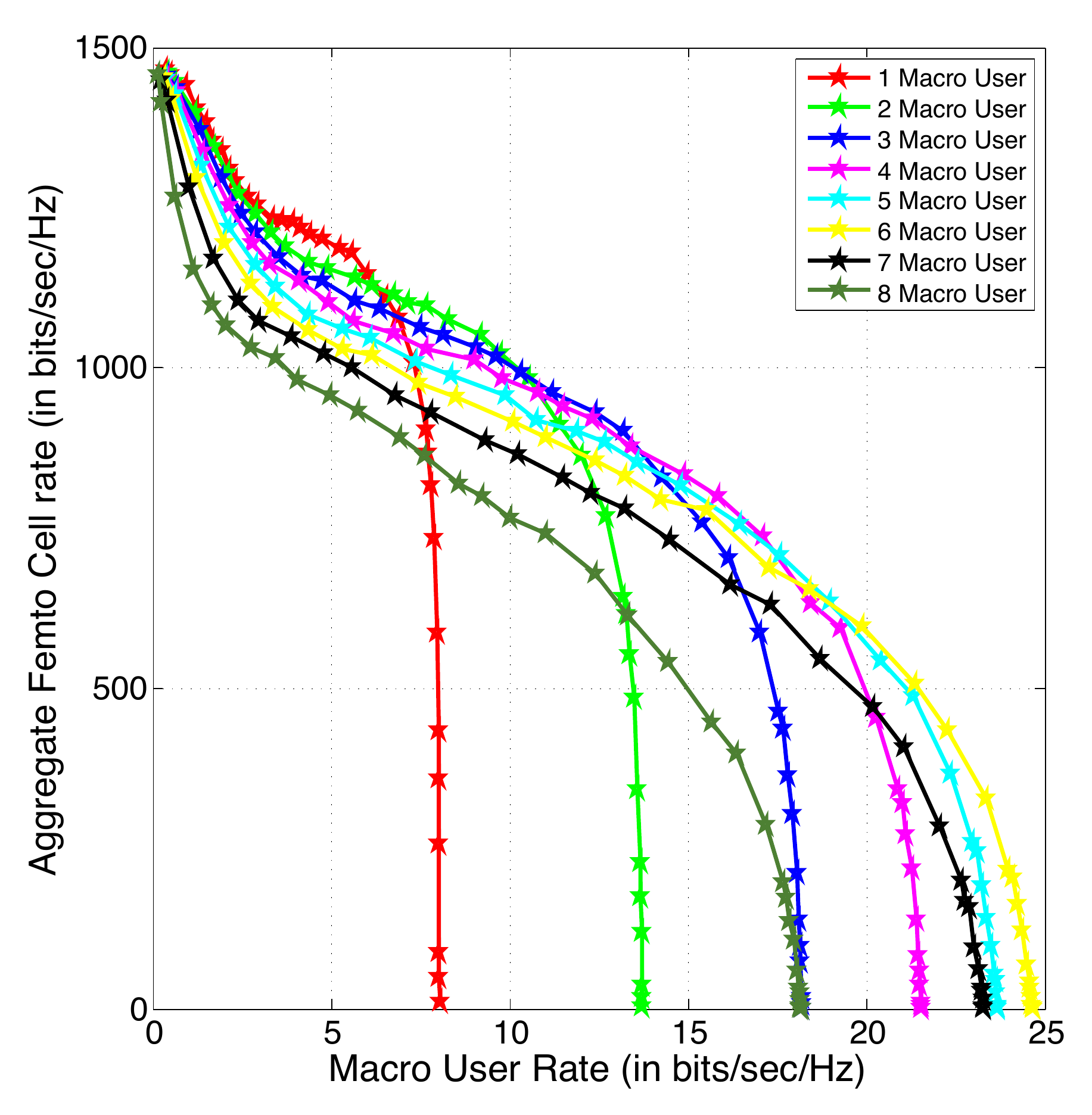}\\
  \caption{Throughput tradeoff region comparison for non-colocated users in macro-DL/femto-UL. 
  The different colors indicate the tradeoff region for different $K$.}\label{fig:non-colocated}
\end{figure}

Figs.~\ref{fig:colocated} and \ref{fig:non-colocated} show 
the throughput tradeoff region (femtocell sum throughput vs. macrocell sum throughput) 
achieved by the proposed scheme in the macro-DL/femto-UL slot
(averaged over random user positions), for 
colocated and non-colocated macro-UTs, respectively.  As the number of served 
macro-UTs increases, the macrocell throughput increases initially
up to a certain maximum value (in our case, the highest macrocell
throughput is obtained for $K = 6$ users) and then decreases. 
This can be expected from the typical behavior of linear LZFB precoding. 
For the non-colocated case, the femtocell throughput decreases significantly as $K$ increases, since
more and more femtocells are affected by the power control interference temperature limitation. This is
because $K$ random macro-UTs positions are selected at each slot. Instead, for colocated macro-UTs, 
the value of $K$ has no effect on the femtocell throughput, therefore, $K = 6$ achieves (approximately) 
uniformly best performance over the whole throughput range. 

\begin{figure}
  \centering
  \includegraphics[width=8cm,height=6cm]{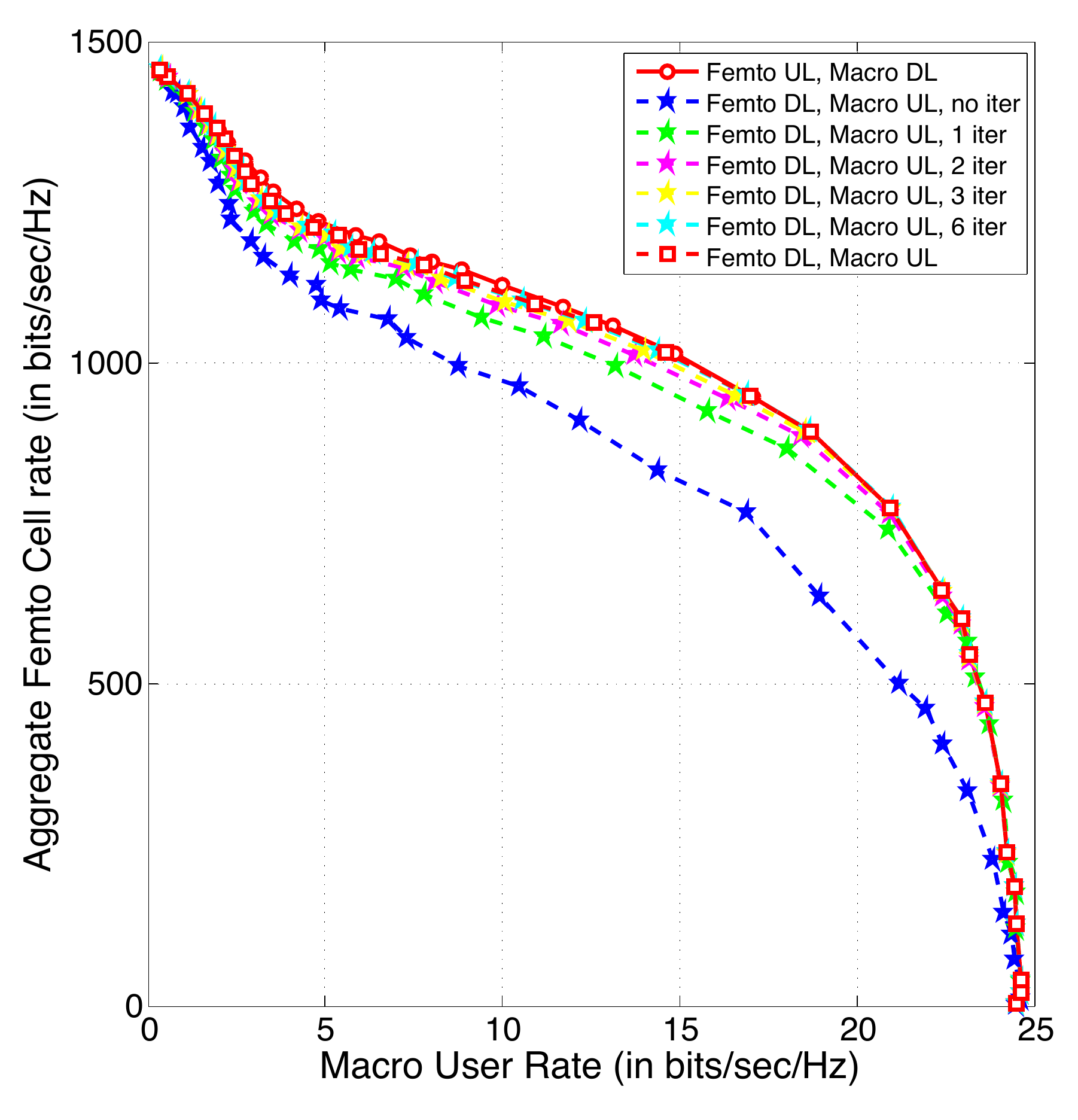}\\
  \caption{Throughput tradeoff region for $K=6$ colocated users in macro-UL/femto-DL for increasing iterations of the power control
  algorithm.}
  \label{fig:colocated-users-6}
\end{figure}

Fig.~\ref{fig:colocated-users-6} and shows the throughput tradeoff 
region for the macro-UL/femto-DL slot, 
for $K = 6$ and different iterations of the power control algorithm. 
In every iteration of the power control algorithm, the peak power
constraint is imposed for both the femtocells as well as the macro
user terminals\footnote{In the macro-DL, each user is given an equal
fraction of the power $P_0$. In the macro-UL, each user is
constrained to use a power no greater than $P_0/K$.}. 
We notice that already for 3 iterations the algorithm yields almost the target ``dual'' macro-DL/femto-UL region, and for 
6 iterations the region is indistinguishable.  Notice also that because of the imposed 
peak power constraint, duality does not strictly hold. 
Nevertheless, as seen from these plots, the impact of the peak power constraint on the macro-UL/femto-DL slot is basically
negligible for this realistic range of system parameters. 

\section{Conclusions}

Overall, the ergodic rate region achievable by
the proposed scheme is very competitive with other schemes proposed
or analyzed in the current literature, considering that it can be
achieved with a very simple protocol and low-complexity signal
processing. 
For example, operating the system at the achievable 
throughput tradeoff point with macrocell throughput of 15 bit/s/Hz and 
femtocell throughput of 1000 bit/s/Hz, we can achieve 600 Mb/s over 40 MHz of system 
bandwidth of  {\em average} macrocell symmetric data rate (both UL and DL), and 
64 Mb/s per femtocell over the same system bandwidth (in our system 
geometry we have 625 femtocells per macrocell).  Given these rather outstanding numbers, 
we believe that the proposed scheme is an attractive option for ``beyond 4G'' future wireless networks.
Of course, several issues need further investigation, as for example the system operations and performance with multiple 
macrocells,  the protocol overhead for implementing cognitive femtocells and adapting the power 
by the iterative algorithm,  and the effect of scheduling co-located macrocell users on the macrocell user channel correlation, 
which may limit the effective macrocell multiplexing gain.

\bibliographystyle{IEEEtran}
\bibliography{references}


\end{document}